\documentclass[10pt,twocolumn,twoside]{IEEEtran}
\usepackage{amssymb}
\usepackage{mathbbol}
\usepackage{graphicx}
\usepackage{amsmath}
\usepackage{array}
\newcommand{\PreserveBackslash}[1]{\let\temp=\\#1\let\\=\temp}
\newcolumntype{C}[1]{>{\PreserveBackslash\centering}p{#1}}
\newcolumntype{R}[1]{>{\PreserveBackslash\raggedleft}p{#1}}
\newcolumntype{L}[1]{>{\PreserveBackslash\raggedright}p{#1}}
\usepackage[usenames]{color}
\usepackage{colortbl,booktabs}
\usepackage{tabularx}
\usepackage{multicol}
\usepackage{booktabs}
\usepackage[Symbol]{upgreek}
\usepackage{subfigure}
\usepackage{bm}
\usepackage{cite}
\usepackage{balance}
\usepackage{threeparttable}
\usepackage{amsmath}
\usepackage{dcolumn}
\usepackage{multirow}
\usepackage{booktabs}
\usepackage{amsfonts}
\newcolumntype{d}[1]{D{.}{.}{#1}}
\usepackage{algorithm} 
\usepackage{algorithmic} 
\usepackage{amsmath}

\usepackage{enumerate}
\usepackage{balance}

\usepackage[switch]{lineno}

\begin{document}

\bibliographystyle{IEEEtran} 

\title{Reconfigurable Intelligent Surface-Based Wireless Communication: Antenna Design, Prototyping and Experimental Results}

\author{\IEEEauthorblockN{Linglong Dai, Bichai Wang, Min Wang, Xue Yang, Jingbo Tan, Shuangkaisheng Bi, Shenheng Xu, Fan Yang, \\ Zhi Chen, Marco Di Renzo, and Lajos Hanzo}
\thanks{L. Dai, B. Wang, M. Wang, X. Yang, J. Tan, S. Bi, S. Xu, and F. Yang are with the Tsinghua National Laboratory for Information Science and Technology (TNList) as well as the Department of Electronic Engineering, Tsinghua University, Beijing 100084, China (E-mail: daill@tsinghua.edu.cn, wbc15@mails.tsinghua.edu.cn, wangmin14@mails.tsinghua.edu.cn, yangxue13@mails.tsinghua.edu.cn, tanjb13@mails.tsinghua.edu.cn, bsks18@mails.tsinghua.edu.cn, shxu@tsinghua.edu.cn, fan\_yang@tsinghua.edu.cn). } %
\thanks{Z. Chen is with the National Key Laboratory of Science and Technology on Communications, University of Electronic Science and Technology of China, Chengdu 611731, China. (E-mail: chenzhi@uestc.edu.cn). } %
\thanks{M. Di Renzo is with the Laboratory of Signals and Systems (CNRS - CentraleSupelec - University of Paris-Sud), Universit¨¦ Paris-Saclay, 91192 Gif-sur-Yvette, France (E-mail: marco.direnzo@l2s.centralesupelec.fr). } %
\thanks{L. Hanzo is with Electronics and Computer Science, University of Southampton, Southampton SO17 1BJ, U.K. (E-mail: lh@ecs.soton.ac.uk). } %
\thanks{This work was supported in part by the National Science and Technology Major Project of China under Grant 2018ZX03001004-003, in part by the National Natural Science Foundation of China for Outstanding Young Scholars under Grant 61722109, in part by the National Natural Science Foundation of China under Grant 61571270, and in part by the Royal Academy of Engineering under the U.K.-China Industry Academia Partnership Programme Scheme under Grant U.K.-CIAPP$\backslash$49.}

}

\maketitle
\begin{abstract}
One of the key enablers of future wireless communications is constituted by massive multiple-input multiple-output (MIMO) systems, which can improve the spectral efficiency by orders of magnitude. However, in existing massive MIMO systems, conventional phased arrays are used for beamforming, which result in excessive power consumption and hardware cost. Recently, reconfigurable intelligent surface (RIS) has been considered as one of the revolutionary technologies to enable energy-efficient and smart wireless communications, which is a two-dimensional structure with a large number of passive elements. In this paper, we propose and develop a new type of high-gain yet low-cost RIS having 256 elements. The proposed RIS combines the functions of phase shift and radiation together on an electromagnetic surface, where positive intrinsic-negative (PIN) diodes are used to realize 2-bit phase shifting for beamforming. Based on this radical design, the world's first wireless communication prototype using RIS having 256 2-bit elements is designed and developed. Specifically, the prototype conceived consists of modular hardware and flexible software, including the hosts for parameter setting and data exchange, the universal software radio peripherals (USRPs) for baseband and radio frequency (RF) signal processing, as well as the RIS for signal transmission and reception. Our performance evaluation confirms the feasibility and efficiency of RISs in future wireless communications. More particularly, it is shown that a 21.7 dBi antenna gain can be obtained by the proposed RIS at 2.3 GHz, while at the millimeter wave (mmWave) frequency, i.e., 28.5 GHz, a 19.1 dBi antenna gain can be achieved. Furthermore, the over-the-air (OTA) test results show that the RIS-based wireless communication prototype developed is capable of significantly reducing the power consumption.
\end{abstract}




\section{Introduction}\label{S1}

\IEEEPARstart MASSIVE multiple-input multiple-output (MIMO) schemes constitute promising techniques for future wireless communications. By relying on a large antenna array, massive MIMO schemes provide a substantial power gain and improve the spectral efficiency by orders of magnitude~\cite{MIMO3}~\cite{MIMO1}. However, in existing massive MIMO systems, where conventional phased arrays are used for beamforming, hundreds of high-resolution phase shifters and complex feeding networks are required~\cite{MIMO2}~\cite{MIMO4}. As a result, the high power consumption and hardware cost of these phase shifters and complex feeding networks limit the antenna array scale in practical massive MIMO systems, which means that the potential advantages of massive MIMO schemes cannot be fully exploited.

Recently, reconfigurable intelligent surfaces (RISs) have been considered as a promising alternative to the traditional phased arrays~\cite{RIS5,RIS1,RIS2,RIS3,RIS4,RRA1}. In contrast to the conventional phased arrays, an RIS consists of a large number of nearly passive elements with ultra-low power consumption, each of which is capable of electronically controlling the phase of the incident electromagnetic waves with unnatural properties, e.g., negative refraction, perfect absorption, and anomalous reflection~\cite{RIS3}~\cite{RRA2}. Moreover, the spatial feeding mechanism of RISs avoids the excessive power loss caused by the bulky feeding networks of phased arrays. Therefore, RISs significantly reduce both the power consumption and hardware cost, although a larger number of antenna elements may be required to guarantee the antenna gain of the conventional phased arrays.

RISs made of low-resolution 1 bit elements have been widely investigated in the literature~\cite{RRA3,RRA4,RRA5,RRA6,RRA7,RRA8,RRA9}. Among them, the current reversal mechanism has attracted extensive attention as a benefit of its phase response, which is near-constant across a wide frequency band~\cite{RRA7,RRA8,RRA9}. However, RISs with 1-bit elements can only provide two phase states, e.g., $0$ and $\pi$. Consequently, they result in more than 3 dB antenna gain reduction due to the significant phase errors~\cite{RRA10}~\cite{RRA11}. To mitigate the performance degradation caused by the 1-bit phase quantization, RISs with multi-bit elements can also be designed at an increased system complexity and hardware cost. It has been shown~\cite{RRA10} that a RIS with 2-bit elements strikes an attractive tradeoff between the performance and complexity, since it has an acceptable antenna gain erosion of about 1 dB caused by the 2-bit phase quantization~\cite{RRA12}. However, only a couple of contributions may be found in the literature on RISs with 2-bit elements~\cite{RRA12}~\cite{RRA13}. Recently, a novel dual linearly/circularly polarized RIS design with 2-bit elements imposing a low magnitude loss has been proposed in our previous 2-page conference report~\cite{RRA14}, where we have designed an electronically controlled RIS with 2-bit elements operating at 1.7 GHz. 

\begin{figure}[tp]
\begin{center}
\includegraphics[width=0.9\linewidth]{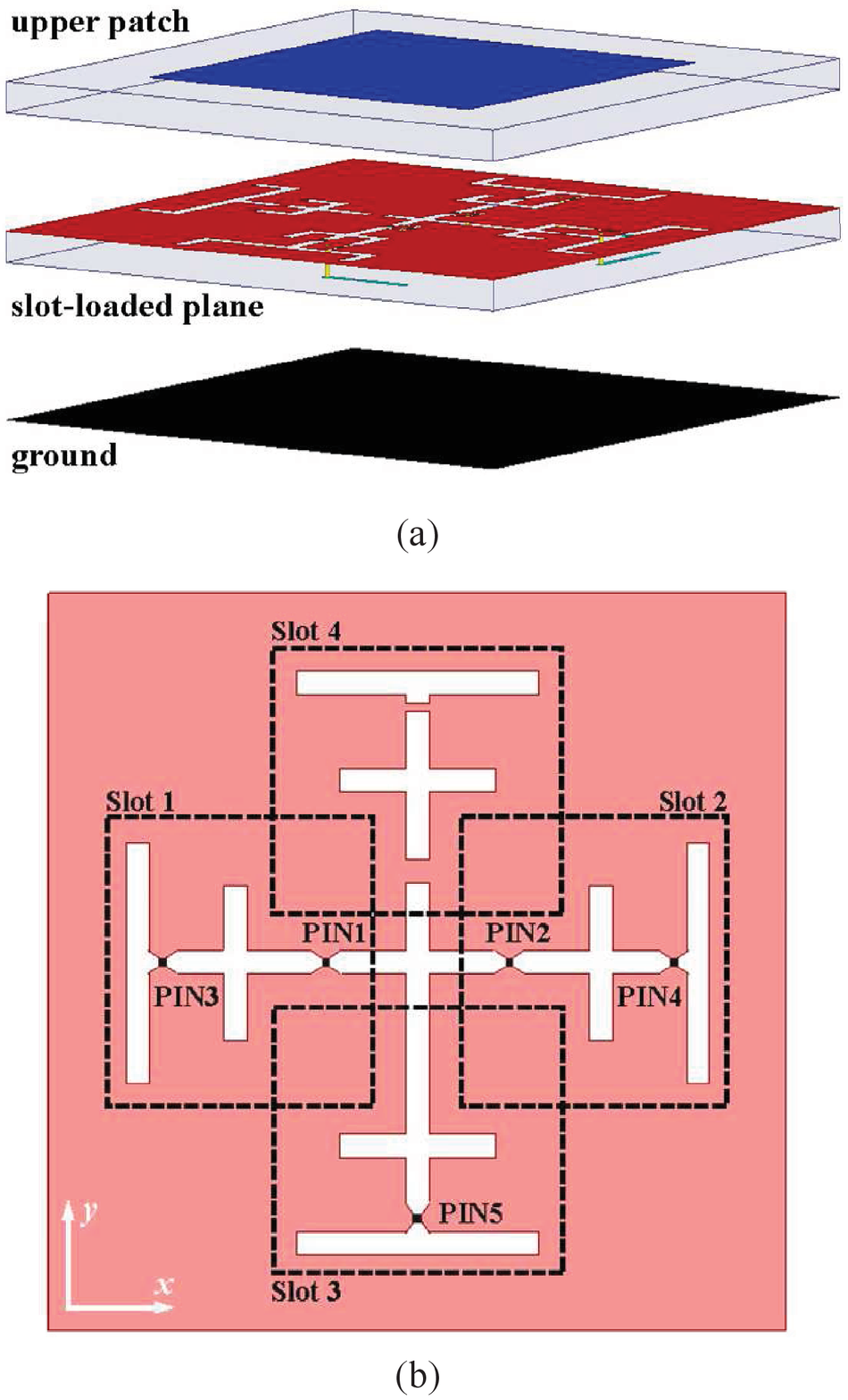} \caption{Structure of the proposed 2-bit RIS element: (a) Exploded view; (b) Detailed view of the slot-loaded plane.} \label{RRA1}
\end{center}
\end{figure}

\begin{figure*}[tp]
\begin{center}
\includegraphics[width=0.8\linewidth]{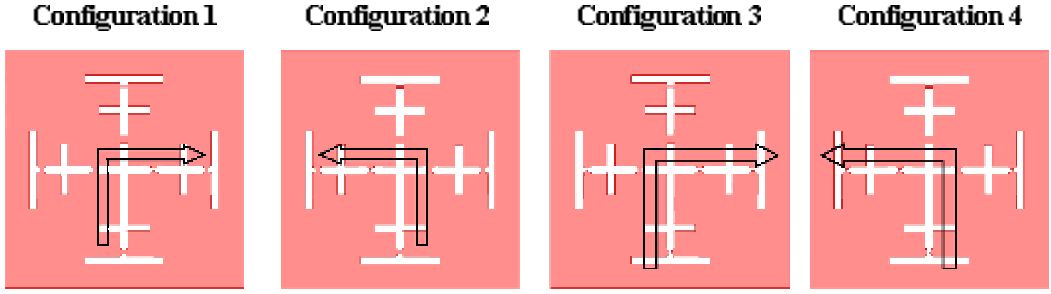} \caption{Illustrations of the RF current paths for four different element configurations.} \label{RRA2}
\end{center}
\end{figure*}

Against this background, we aim for realizing energy-efficient wireless communications by using RISs instead of conventional phased arrays. More particularly, we fabricate and measure electronically controlled RISs with 2-bit elements at 2.3 GHz and 28.5 GHz having $16 \times 16$ elements for the first time. Explicitly, we design the world's first wireless communication prototype using a RIS having 256 2-bit elements\footnote{In this paper, the RIS is deployed at the transmitter~\cite{RIS2} for performance evaluation, which can be also used for intelligent reflection relay~\cite{RIS1}~\cite{RIS2}.}. It should be noted that a programmable metasurface-based wireless communication prototype has been developed in~\cite{RIS4}, where the metasurface is used to modulate the signals for transmission only, while the RIS is used for beamforming both for transmssion and reception in this paper. Specifically, the prototype designed consists of modular hardware and flexible software to realize the wireless transceiver functions, including the hosts for parameter setting and data exchange, the universal software radio peripherals (USRPs) for baseband and radio frequency (RF) signal processing, as well as the RIS for signal transmission and reception. More particularly, the USRP at the transmitter firstly performs baseband signal processing, e.g., source coding, channel coding and orthogonal frequency division multiplexing (OFDM) modulation. Then, the RF signals output by the RF chains are transmitted via our RIS having 256 2-bit elements\footnote{In this paper, a single RF chain is considered, which can be easily extended to multiple RF chains.}. The contaminated signals are received by the receiver antenna, and then the USRP at the receiver takes charge of both RF and baseband signal processing for recovering the original signals. Our performance evaluation confirms the feasibility and efficiency of RISs in wireless communication systems. More specifically, it is shown that a 21.7 dBi antenna gain can be obtained by the proposed RIS at 2.3 GHz, while at 28.5 GHz, a 19.1 dBi antenna gain can be achieved. Furthermore, the over-the-air (OTA) test results show that the RIS-based wireless communication prototype developed significantly reduces the power consumption, while achieving similar or better performance in terms of effective isotropic radiated power (EIRP), compared to conventional phased array-based wireless communications.

The rest of this paper is organized as follows. The basic principles and implementation details of the RIS having 256 2-bit elements are introduced in Section II. The RIS-based wireless communication prototype designed is described in Section III. Section IV shows the experimental results. Finally, our conclusions are drawn in Section V.

\section{The RIS with 2-Bit Elements}\label{S2}

For high-quality antenna array design, it is crucial to accurately control the aperture field, especially the phase distribution for electronically forming a sharp pencil beam. The conventional phased array utilizes a phase shifter connected to each antenna element in the array for directly controlling the element's complex excitation with a specific phase state. The massive number of phase shifters required becomes the primary contributing factor to the high power consumption and excessive hardware cost of conventional phased arrays. By contrast, the proposed RIS with 2-bit elements simply employs positive intrinsic-negative (PIN) diodes integrated in each element, which modulate the RF currents induced in the antenna elements upon their illumination by turning ON or OFF the PIN diodes. Therefore, the RIS element becomes capable of re-radiating an electromagnetic field having a specific phase state, hence realizing the desired electronic phase control capability without using conventional phase shifters.

The most essential breakthrough in the proposed RIS with 2-bit elements is its novel antenna element structure. As depicted in Fig. \ref{RRA1} (a), each antenna element consists of a square-shaped upper patch, a slot-loaded plane and a ground plane. The upper patch receives and radiates energy, while the ground plane suppresses the back radiation and radiates through the slots in the slot-loaded plane. The slot-loaded plane is the key component controlling the RIS's phase state, and its detailed structure is shown in Fig. \ref{RRA1} (b). Four sets of slots are positioned symmetrically, on which five PIN diodes are integrated.

Ideally, a 2-bit RIS element provides four quantized phase states with a $90^\circ$ phase increment. The PIN diode states are appropriately combined for beneficially controlling the RF current paths and the resonant lengths of the slots, thus resulting in tunable phase shifts for the proposed RIS element design. Specifically, Slots 1, 2 and 3 are arranged in a T-shaped configuration, which are complemented by a dummy Slot 4 invoked for maintaining a symmetric element structure. The states of PIN 1 and PIN 2 are alternatively turned ON or OFF, so that the currents induced may be reversed in the $x$-direction, corresponding to a $180^\circ$ phase shift. Furthermore, the other 3 PIN diodes are turned ON or OFF simultaneously for the sake of changing the resonant lengths of the slots, so that an additional $90^\circ$ phase shift may be realized. Hence, a 2-bit phase resolution can be obtained, resulting in four different phase states.

\begin{table}[tp]
\begin{center}
\caption{PIN diode states for different element configurations.} \label{t1}
\begin{tabular}{|c|c|c|c|c|}
\hline
\textbf{Configuration} & \textbf{PIN1/PIN2} & \textbf{PIN3} & \textbf{PIN4} & \textbf{PIN5}\\
\hline
1 & ON/OFF & \multicolumn{3}{c|}{\multirow{2}{*}{ON}} \\
\cline{1-2}
2 & OFF/ON & \multicolumn{3}{c|}{\multirow{2}{*}{}} \\
\hline
3 & ON/OFF & \multicolumn{3}{c|}{\multirow{2}{*}{OFF}} \\
\cline{1-2}
4 & OFF/ON & \multicolumn{3}{c|}{\multirow{2}{*}{}} \\
\hline
\end{tabular}
\end{center}
\end{table}

The PIN diode states of the four element configurations are tabulated in TABLE I, and a graphic illustration of the RF current paths is presented in Fig. \ref{RRA2}. It is worthwhile pointing out that the incident and re-radiated fields of each element are orthogonally polarized due to the change of current directions. Thanks to the symmetric element structure, the proposed RIS element is capable of providing the same performance both under an $x$- and a $y$-polarized incident wave, hence it is suitable for both dual-linearly and dual-circularly polarized systems.

The proposed 2-bit RIS element operates in the S band having a center frequency of 2.3 GHz\footnote{For simplicity, we only provide details about the RIS at 2.3 GHz, since the design philosophy at 28.5 GHz is the same.}. The element spacing is 50 mm. The upper patch has a size of $37 \times 37$ mm$^2$ and it is etched on a 1-mm thick FR4 substrate. The slot-loaded plane is etched on another FR4 substrate, which is placed 6 mm below the upper patch. The ground is made of an aluminum sheet, placed 12 mm below the slot-loaded plane.

\begin{figure}[tp]
\begin{center}
\includegraphics[width=1\linewidth]{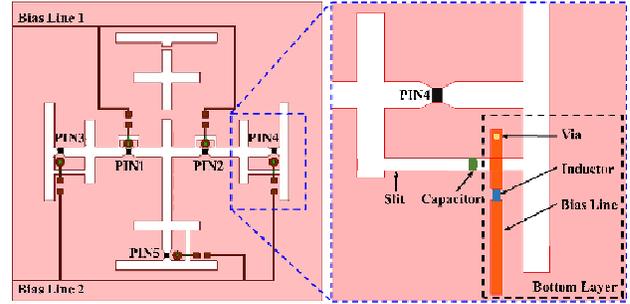} \caption{Layout of the DC bias network of the 2-bit RIS element.} \label{RRA3}
\end{center}
\end{figure}

The type of the commercial PIN diode is SMP1340-040LF from Skyworks. The ON state of the PIN diode is modeled as a series of $R = 0.8 \ \Omega$ lumped resistors and $L = 780$ pH inductors. The OFF state can be modeled as a series of $C = 202$ fF lumped capacitors, $R = 10 \ \Omega$ resistors and $L = 780$ pH inductors. The DC bias network of the element is depicted in Fig. \ref{RRA3}. The five PIN diodes are divided into two groups, which are then independently controlled by a pair of bias lines. The forward and reverse DC voltages are $+0.9$ V and $-0.9$ V, respectively.

\begin{figure}[tp]
\centering
\subfigure[]{
\begin{minipage}{8cm}
\centering
\includegraphics[width=1.05\linewidth]{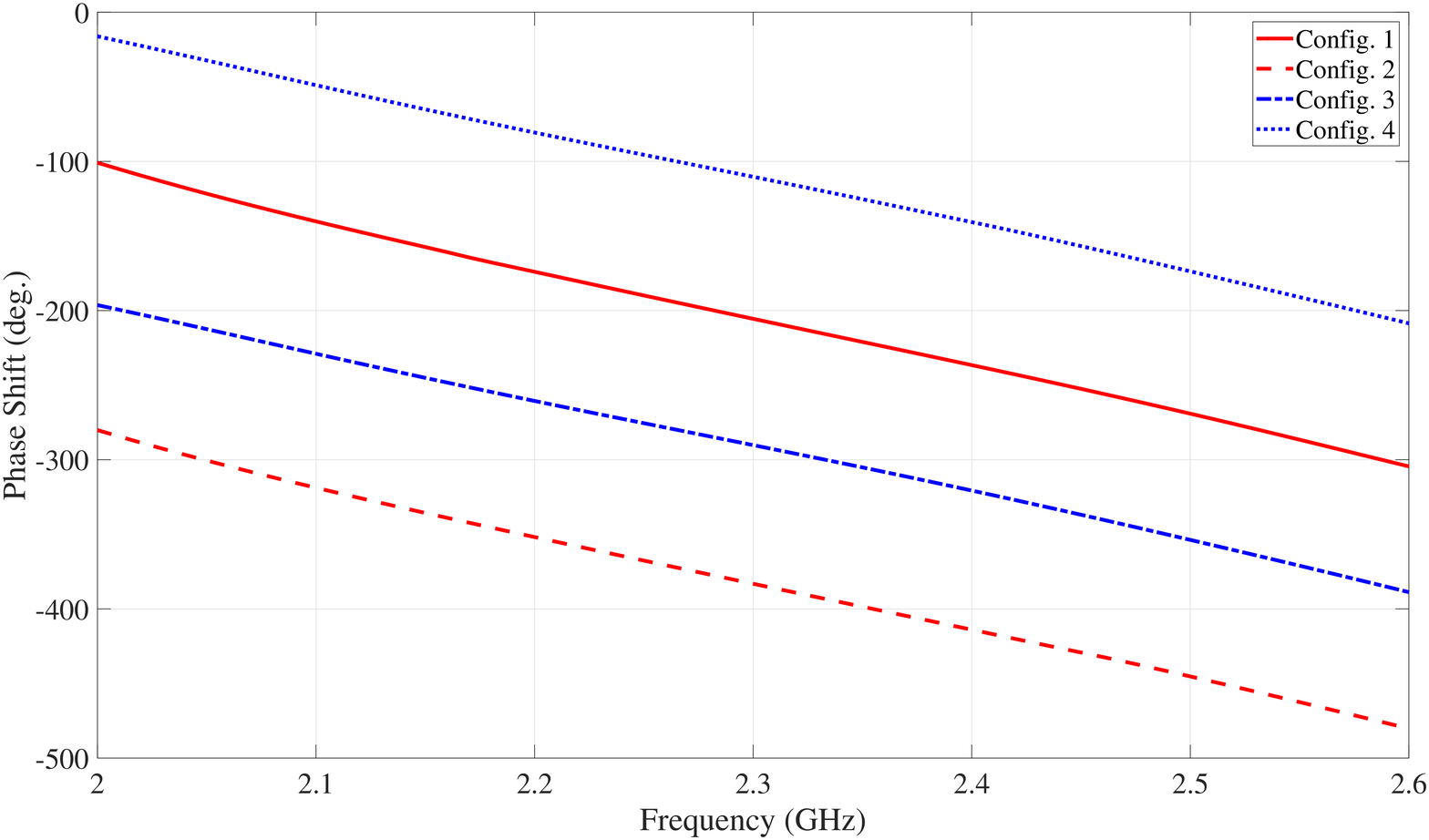}
\end{minipage}
}

\subfigure[]{
\begin{minipage}{8cm}
\centering
\includegraphics[width=1.05\linewidth]{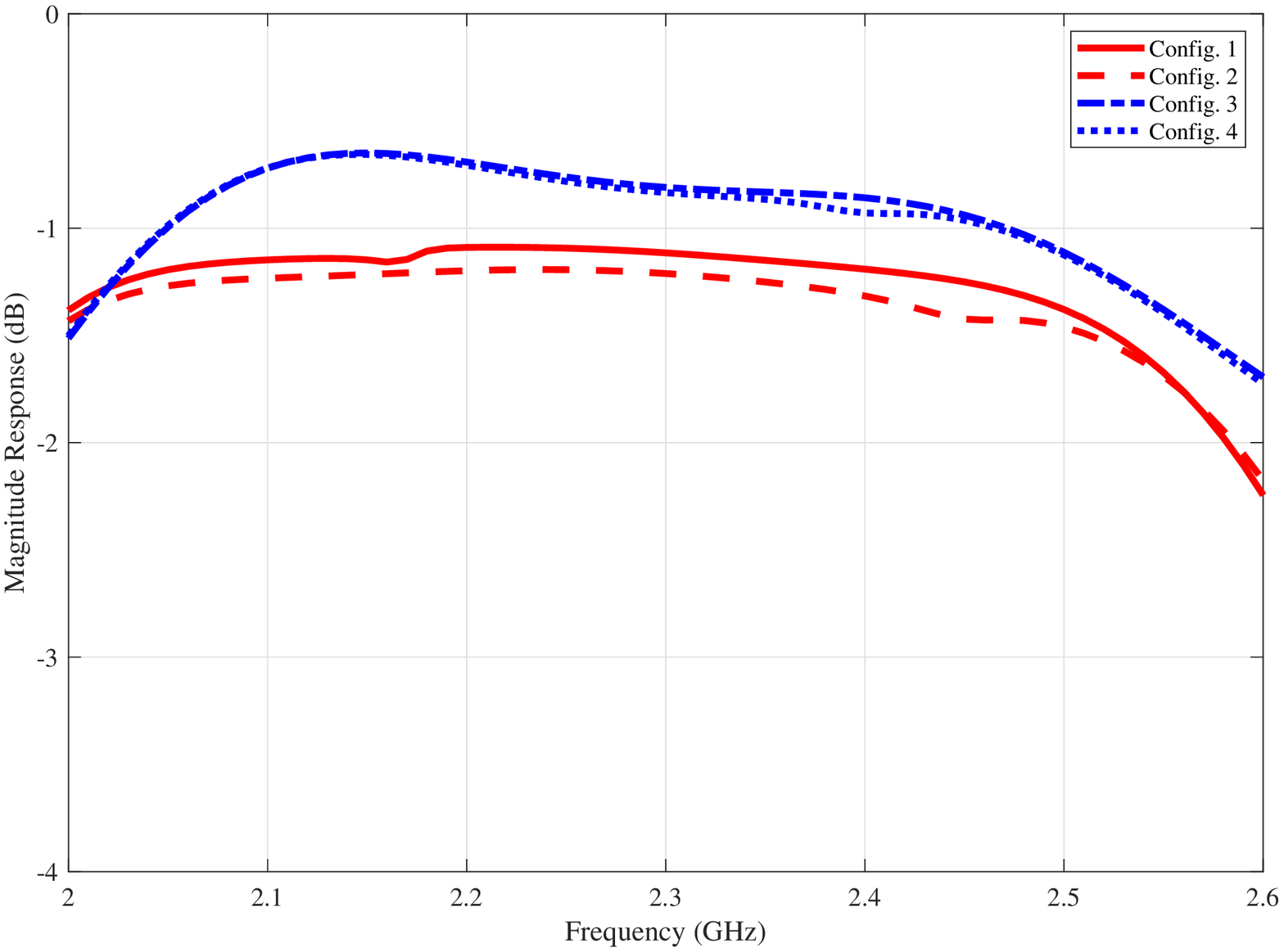}
\end{minipage}
}
\caption{Simulated performances of the proposed 2-bit RIS element: (a) Phase performance; (b) Magnitude performance.} \label{RRA4}
\end{figure}

\begin{table}[tp]
\begin{center}
\caption{Simulated element performance at 2.3 GHZ.} \label{t2}
\begin{tabular}{|c|c|c|}
\hline
\textbf{Configuration} & \textbf{Phase Shift} & \textbf{Magnitude Response}\\
\hline
1 & $-205.5^\circ$ & $-1.1$ dB \\
\hline
2 & $-383.2^\circ$ & $-1.2$ dB \\
\hline
3 & $-290.2^\circ$ & $-0.8$ dB \\
\hline
4 & $-110.3^\circ$ & $-0.8$ dB \\
\hline
\end{tabular}
\end{center}
\end{table}

\begin{figure}[tp]
\begin{center}
\includegraphics[width=0.8\linewidth]{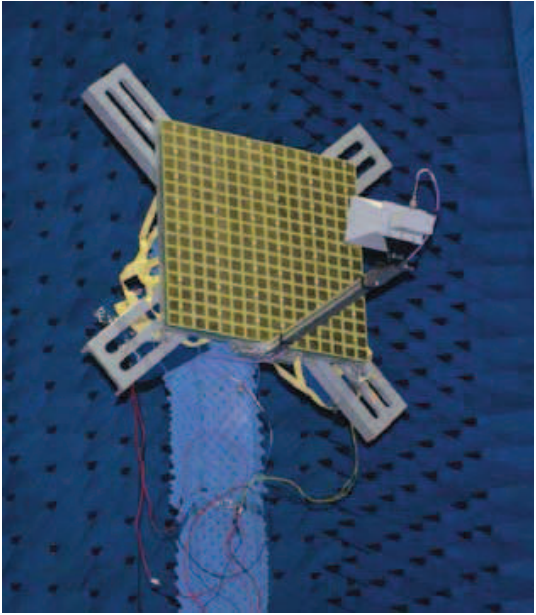} \caption{Photograph of the fabricated RIS having $16 \times 16$ 2-bit elements.} \label{RRA5}
\end{center}
\end{figure}

\begin{figure}[tp]
\begin{center}
\includegraphics[width=0.9\linewidth]{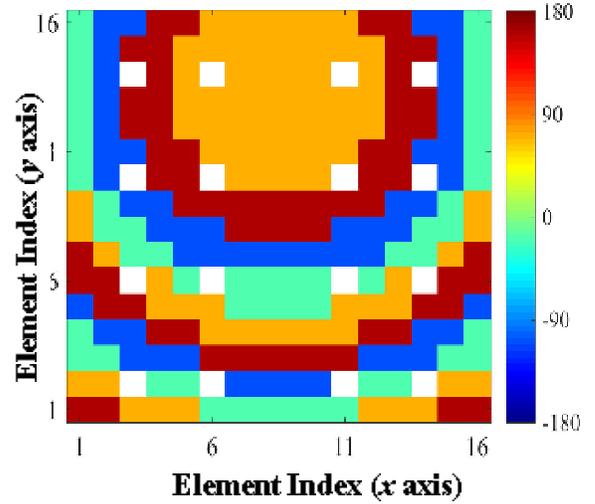} \caption{Phase shift distribution of the RIS elements for the broadside beam. Note that 16 elements (labeled with white color) are removed for wiring the bias lines.} \label{RRA6}
\end{center}
\end{figure}

The simulated phase and magnitude performances of the proposed 2-bit RIS element for four element configurations are shown in Fig. \ref{RRA4}, and the exact values at 2.3 GHz are summarized in TABLE II. It can be observed that the four element phase states clearly exhibit a 2-bit phase resolution with an approximate phase increment of $90^\circ$. These states remain very stable within the frequency band of interest spanning from 2 GHz to 2.6 GHz. The insertion magnitude loss is less than 1.2 dB, which slightly deteriorates at higher frequencies above 2.5 GHz. Because of the current reversal mechanism, the magnitude responses of the element configurations 1 and 2 (or 3 and 4) are similar, while their phase shift difference is around $180^\circ$. These simulated results successfully demonstrate the electronic phase shifting capability of the proposed 2-bit RIS element without using phase shifters.

Moreover, a RIS having $16 \times 16$ 2-bit elements is designed and fabricated, as shown in Fig. \ref{RRA5}. The size of the surface is 800 mm $\times$ 800 mm, and the distance between the primary feed and the surface is 720 mm. Upon being illuminated by the primary feed, these $16 \times 16$ elements can be dynamically reconfigured to convert the spherical wavefront impinging from the feed into a planar wavefront in the desired direction. Hence, the RIS designed is capable of producing a focused high-gain beam that is capable of promptly switching its direction within a two-dimensional $\pm 60^\circ$ angular range. As an example, Fig. \ref{RRA6} shows the phase shift distribution of the elements for the broadside beam, where 16 elements (labeled with white color) are removed for wiring the bias lines. For practical realization, an FPGA-based beamforming control board is designed, which provides 256 DC bias signals for individually setting the configuration of all $16 \times 16$ elements. We can also form a large variety of shaped beams provided that the appropriate configurations of the elements are determined using a phase-only synthesis process and are then pre-loaded into the beamforming control board.

\begin{figure}[tp]
\begin{center}
\includegraphics[width=1\linewidth]{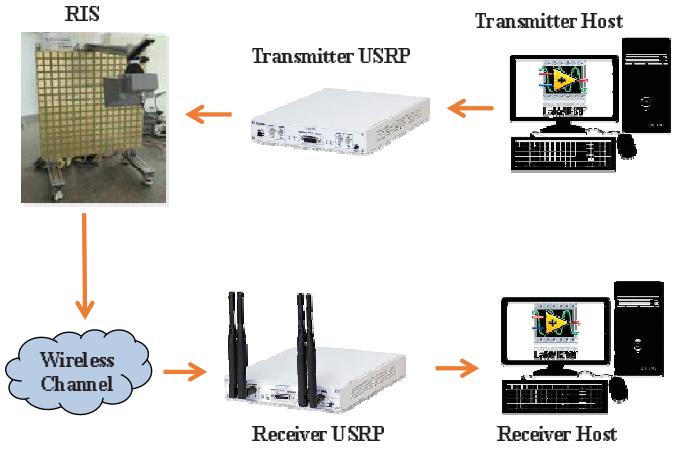} \caption{The RIS-based wireless communication prototype.} \label{PRO1}
\end{center}
\end{figure}

\begin{figure}[tp]
\begin{center}
\includegraphics[width=1\linewidth]{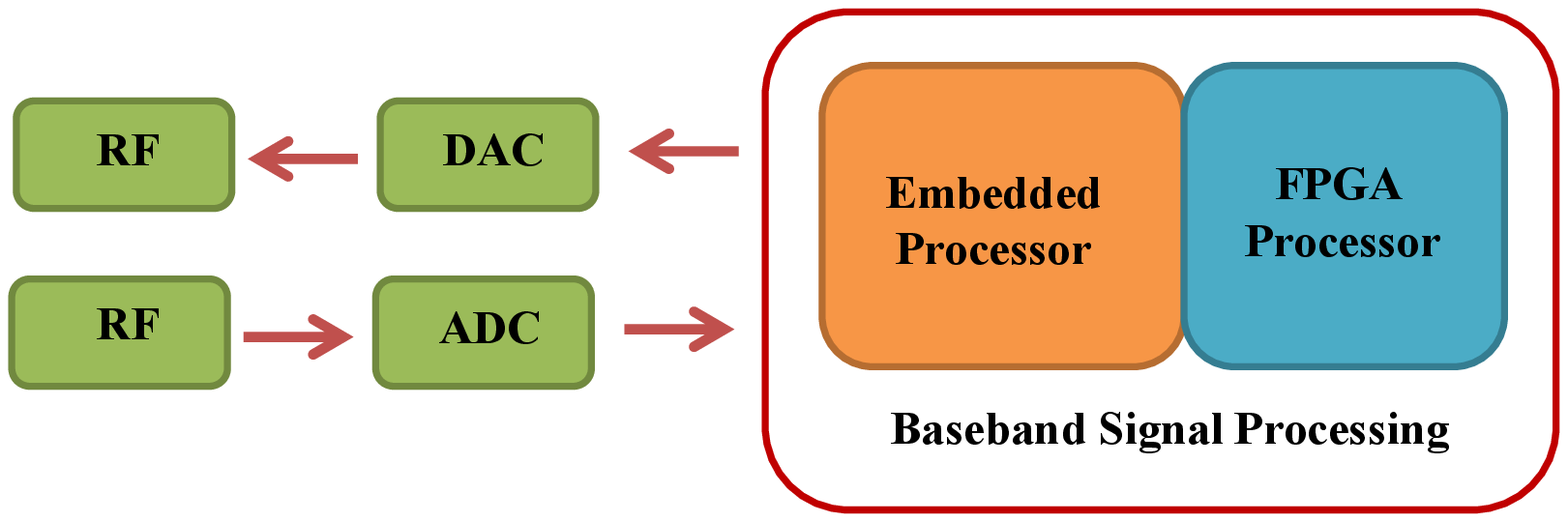} \caption{The hardware modules of USRP.} \label{PRO2}
\end{center}
\end{figure}

\section{RIS-Based Wireless Communication Prototype}\label{S3}

\begin{figure*}[tp]
\begin{center}
\includegraphics[width=0.9\linewidth]{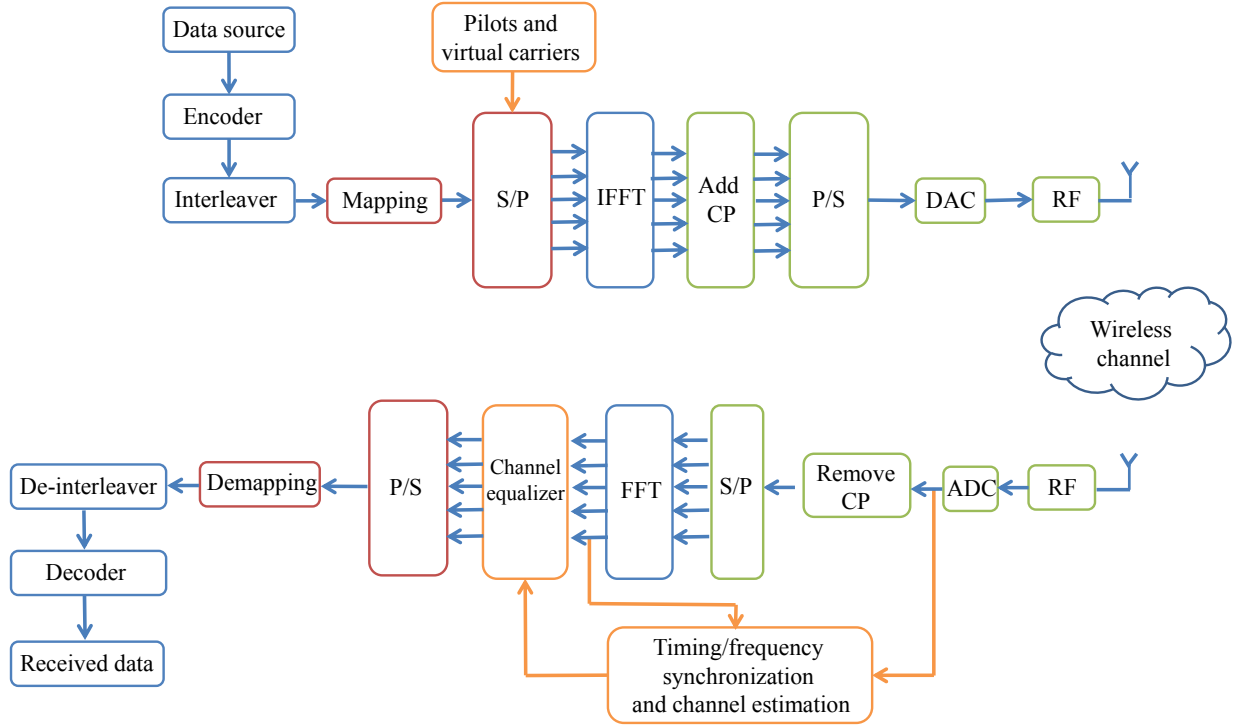} \caption{Signal processing flow.} \label{PRO3}
\end{center}
\end{figure*}

The RIS-based wireless communication prototype designed consists of modular hardware and flexible software, which collectively realize our end-to-end wireless communication system, including baseband signal processing, RF transmission, etc.

As shown in Fig. \ref{PRO1}, the hardware structure of the RIS-based wireless communication prototype designed consists of the base station side, including the transmitter host, the USRP at the transmitter and the RIS having 256 2-bit elements, as well as the user side, including the receiver antenna, the USRP at the receiver and the receiver host.

At the base station side, a graphical interface is realized at the transmitter host, which is responsible for controlling the parameters at the transmitter, including the carrier frequency, transmit power, modulation and encoding modes, etc. Then, the signals are delivered to the USRP at the transmitter via the transmitter host. Once the signals from the transmitter host have been received, the USRP at the transmitter carries out the signal processing, which aims to transform the signals into the form suitable for transmission in wireless channels. The detailed signal processing flow will be introduced later in this section. After that, the processed signals are forwarded to the RIS having 256 2-bit elements. As discussed in Section II, by adjusting the states of the PIN diodes to control the phase for each element, a sharp directional beam can be generated by the RIS for transmission to the user side.

At the user side, the contaminated signals are received from the wireless channel, which are then processed by the USRP at the receiver. The USRP at the receiver takes charge of the signal processing for recovering the original signals, which is basically the inverse process of those in the USRP at the transmitter. Finally, the recovered signals and the corresponding parameters, such as the received signal power, constellation, bit-error-rate (BER), data rates and so on, are displayed by the graphical interface at the receiver host.

To elaborate a little further, observe in Fig. \ref{PRO2} that the USRP at the transmitter consists of four modules: the embedded processor, the high-speed FPGA processor, the digital-to-analog (DA) module and the RF module. To be more specific, the embedded processor and the high-speed FPGA processor jointly realize the baseband signal processing. Specifically, the embedded processor handles the media access control (MAC) layer process, such as data framing, while the high-speed FPGA module realizes the physical layer signal processing, such as channel coding and orthogonal frequency division multiplexing (OFDM) modulation. The DA module is used for the digital-to-analog conversion (DAC) of the digital signals output by the FPGA processor, and the RF module realizes the up-conversion required for RF signal transmission. Similarly, the USRP at the receiver also consists of four modules: the embedded processor, the high-speed FPGA processor, the analog-to-digital (AD) module and the RF module. The baseband signal processing at the receiver is the inverse procedure of that at the transmitter, while the AD module is used for analog-to-digital conversion (ADC) of the analog signals obtained by down-conversion. Finally, the down-conversion of the received RF signal is realized by the RF module.

\begin{figure}[tp]
\begin{center}
\includegraphics[width=0.9\linewidth]{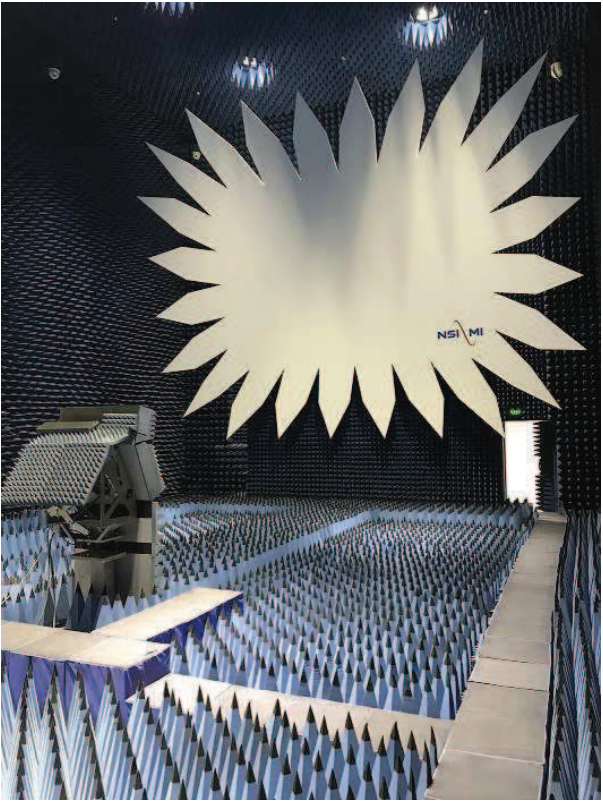} \caption{The compact range anechoic chamber of size $20 \times 10 \times 10$ $\rm{m}^3$.} \label{RRA9}
\end{center}
\end{figure}

In our prototype, the baseband signal processing procedure follows the LTE standard relying on frequency division duplex (FDD)~\cite{LTE1,LTE2,LTE3}, and is implemented by a high-speed FPGA processor as part of the USRP with the aid of graphical programming.

Specifically, the signal processing flow of the system is shown in Fig. \ref{PRO3}. The signals can be obtained from diverse data sources, such as text, images, videos and so on. To achieve efficient and robust wireless communications, a series of signal precessing operations have to be carried out.

Firstly, the input signals are transferred to the encoder, including source encoding and channel encoding, where source encoding is performed to reduce redundancy inherent in the multimedia input signals, and thus can realize more efficient transmission. By contrast, channel coding aims for combatting the channel-induced impairments by correcting the transmission errors. After that, bit-interleaving is applied for dispersing the burst errors into random errors, hence improving the channel coding performance, especially for channels with memory. Then, the interleaved bits are mapped to symbols according to the modulation modes, such as phase shift keying (PSK) or quadrature amplitude modulation (QAM), where the different modulation modes will result in different data rates, depending on the number of bits/symbol.

Furthermore, as seen in Fig. \ref{PRO3}, OFDM is adopted for wideband transmission over dispersive channels. In this regard, after adding the pilots and virtual subcarriers, the serial stream of symbols is serial-to-parallel converted and mapped to the frequency-domain OFDM subcarriers. Then, the frequency-domain OFDM symbols are transformed to the time-domain by the inverse fast fourier transform (IFFT). After concatenating the cyclic prefix (CP), the OFDM symbols are converted to the serially transmitted time-domain signals. Afterwards, the serially transmitted signals are forwarded to the DAC module and to the up-conversion module, and finally are transmitted via the RIS.

The receiver basically carries out the inverse process of the transmitter, where the synchronization signals and the pilots are utilized for timing/frequency synchronization and channel estimation. Finally, the estimated channel will be used for signal detection.

The above signal processing flow is controlled by the software system. By changing the system parameters, such as the transmit power, coding modes, modulation modes, etc., various transmission modes can be activated according to specific scenarios and requirements.

\begin{figure}[tp]
\centering
\subfigure[]{
\begin{minipage}{8cm}
\centering
\includegraphics[width=1.05\linewidth]{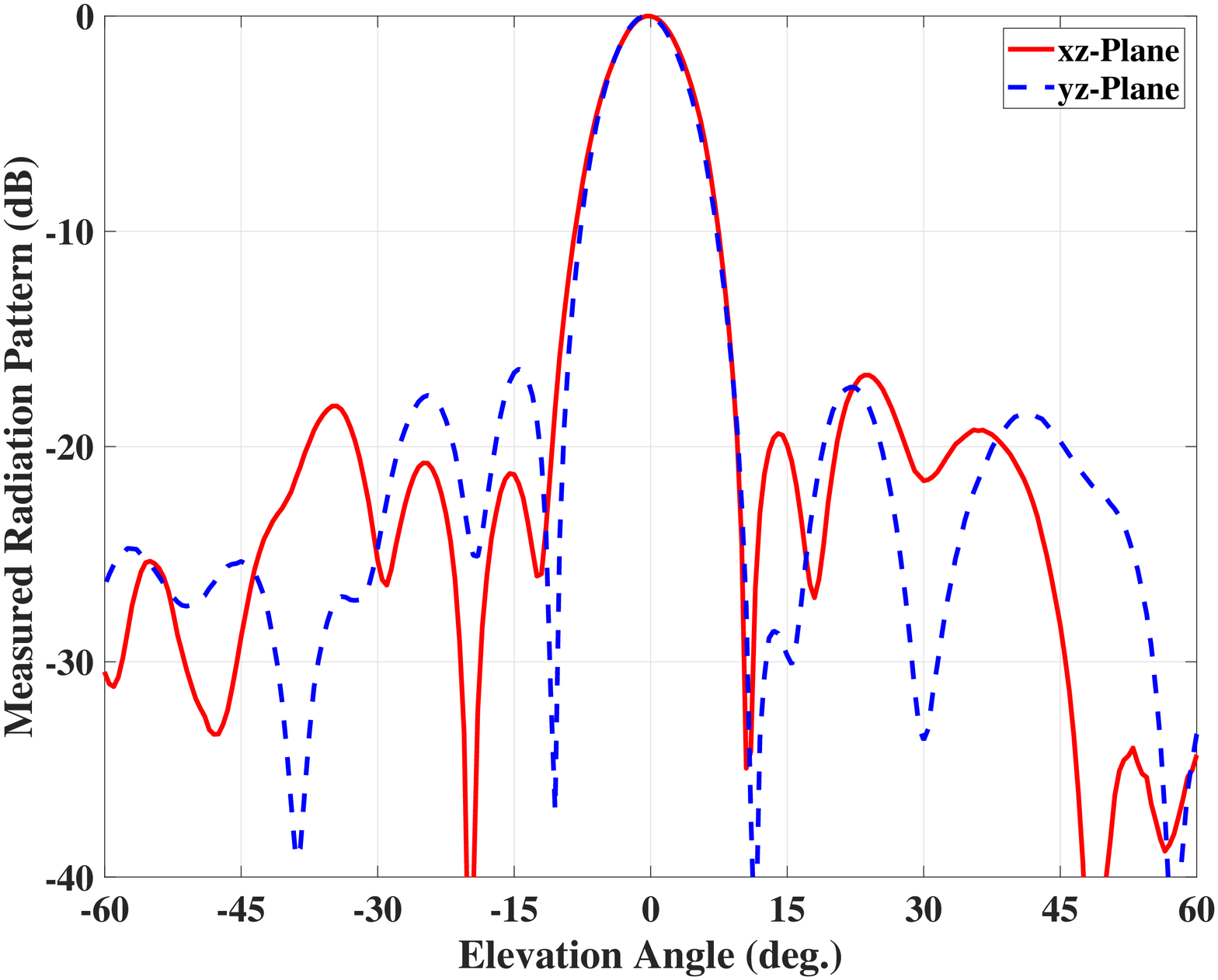}
\end{minipage}
}

\vspace*{+1mm}

\subfigure[]{
\begin{minipage}{8cm}
\centering
\includegraphics[width=1.05\linewidth]{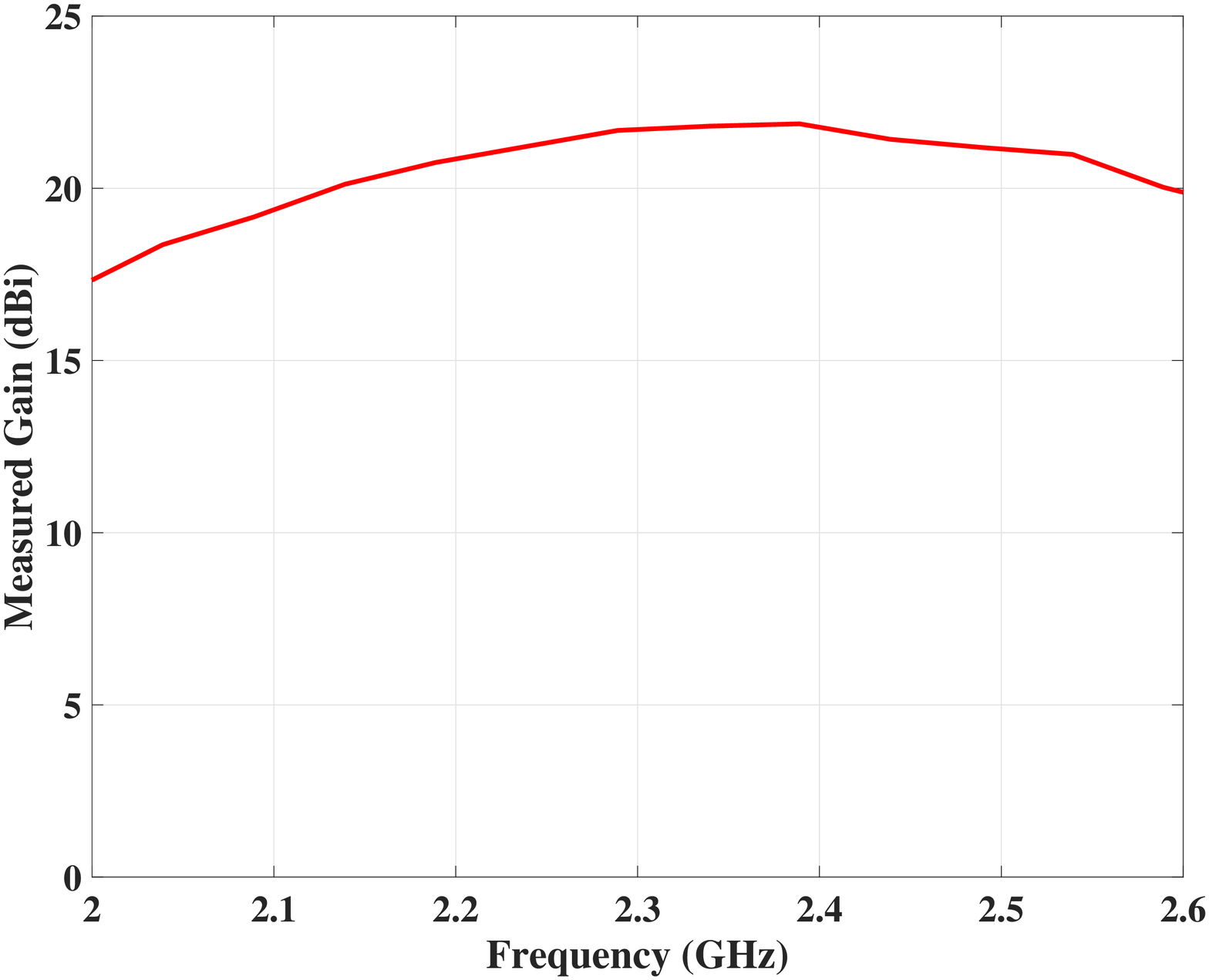}
\end{minipage}
}
\caption{Measured performance of the broadside beam: (a) Normalized radiation patterns at 2.3 GHz; (b) Antenna gains within the frequency band of interest (from 2 GHz to 2.6 GHz).} \label{RRA7}
\end{figure}

\begin{figure}[tp]
\begin{center}
\includegraphics[width=1\linewidth]{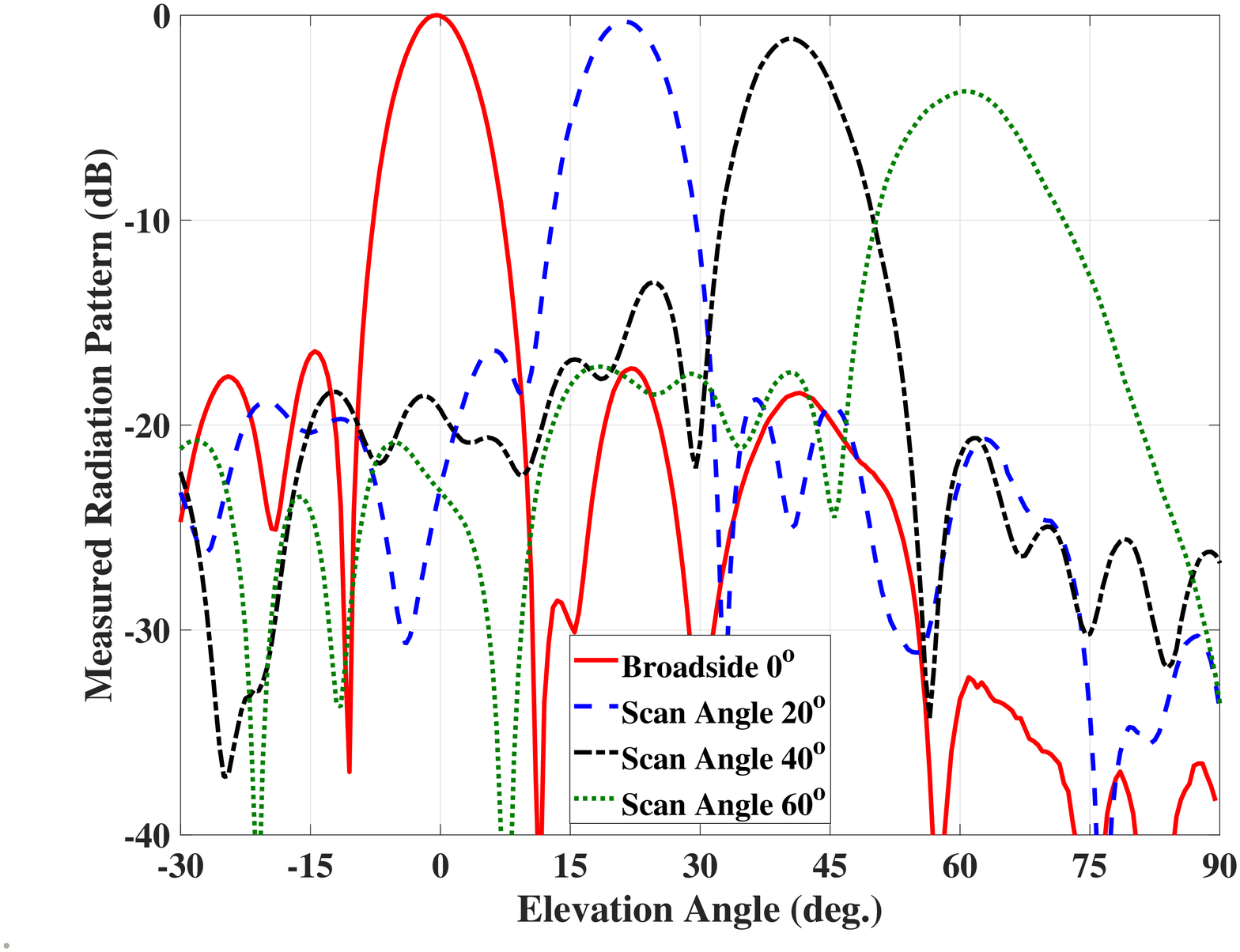} \caption{Measured radiation patterns of the scanned beams. They are normalized to the measured gain of the broadside beam.} \label{RRA8}
\end{center}
\end{figure}

\section{Experimental Results}\label{S4}

The RIS with 2-bit elements constructed is measured using a compact anechoic chamber as shown in Fig. \ref{RRA9}, which is built in Tsinghua University, and the room size is $20 \times 10 \times 10$ $\rm{m}^3$. Fig. \ref{RRA7} (a) shows the normalized measured radiation patterns of the broadside beam at the design frequency of 2.3 GHz. A high-gain pencil beam is formed by controlling the phase shifts of the 2-bit RIS elements. The half-power beamwidths are $9.1^\circ$ and $8.8^\circ$ in the two principal planes, respectively, and the measured sidelobe levels are $-16.7$ dB and $-16.4$ dB. The measured antenna gains within the frequency band of interest are plotted in Fig. \ref{RRA7} (b). At 2.3 GHz, the measured gain is 21.7 dBi, which corresponds to an aperture efficiency of $31.3\%$. The measured gain reaches its maximum of 21.9 dBi at 2.4 GHz, and the 1-dB gain bandwidth is 350 MHz, which is equivalent to $15.2\%$ at the design frequency of 2.3 GHz.

The beams scanned from $0^\circ$ to $60^\circ$ can be readily obtained with the help of our beamforming control board. The measured radiation patterns, normalized to the gain of the broadside beam, are plotted in Fig. \ref{RRA8}. As the scanning angle increases, the measured gain decreases and the main beam broadens. When the scanning angle is $60^\circ$, the measured gain is 18.0 dBi, and the scanning gain reduction is only 3.7 dB. These measured results successfully verify the flexible wide-angle beam-scanning capability of the proposed RIS with 2-bit elements.

\begin{figure}[tp]
\begin{center}
\includegraphics[width=1\linewidth]{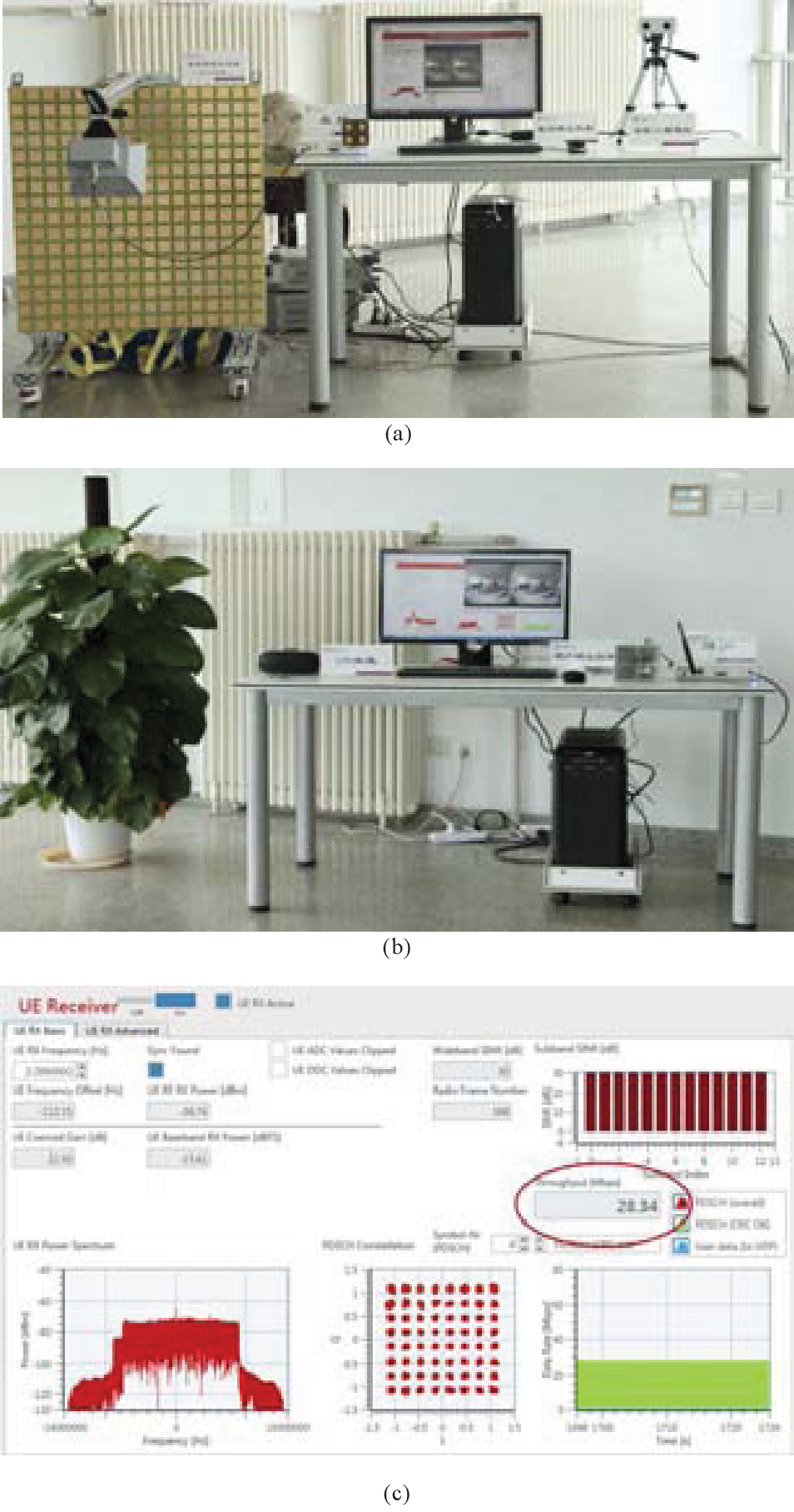} \caption{The constructed RIS-based prototype at 2.3 GHz: (a) Transmitter; (b) Receiver; (c) Graphical interface at the receiver side.} \label{PRO4}
\end{center}
\end{figure}

Based on the proposed RIS with 2-bit elements, we construct the transmitter and receiver of the RIS-based wireless communication prototype shown in Fig. \ref{PRO4} (a) and Fig. \ref{PRO4} (b), respectively. The OTA test environment is indoor, and the distance between the transmitter and receiver is 20 meters. High-definition virtual reality (VR) video streams captured by a stereoscopic camera is used as the data source in the RIS-based wireless communication prototype for real-time demonstration.

\begin{figure}[tp]
\begin{center}
\includegraphics[width=0.9\linewidth]{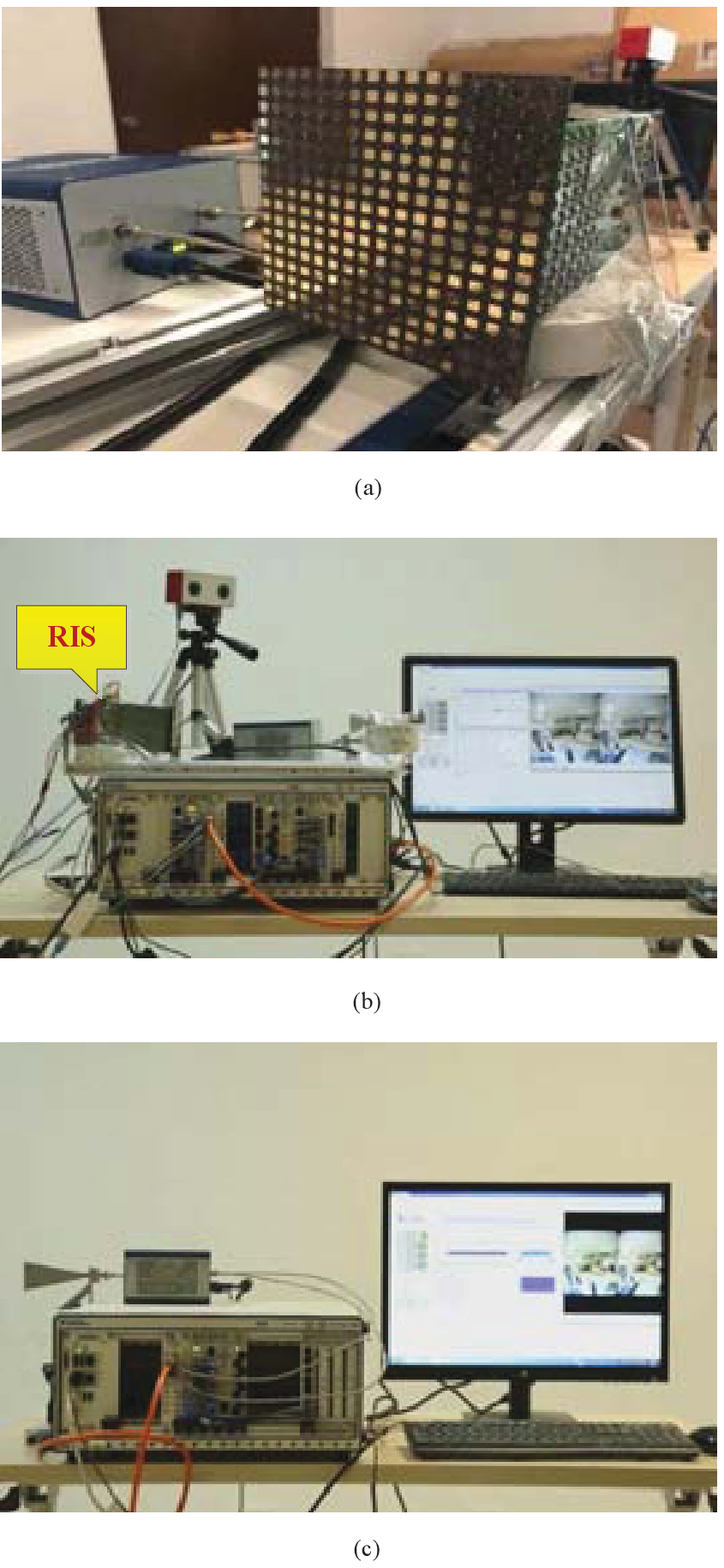} \caption{The constructed RIS-based prototype at 28.5 GHz: (a) The RIS operating at 28.5 GHz; (b) Transmitter; (c) Receiver.} \label{PRO5}
\end{center}
\end{figure}

The parameters of the test are listed as follows: 1) The resolution of the VR camera is 3920 $\times$ 1440, and the frame rate is 30; 2) The carrier frequency of OFDM modulation is 2.3 GHz, and the number of subcarriers is 1200; 3) A variety of modulation modes are available, including QPSK, 16QAM, 64QAM, etc, and Turbo encoding is used in conjunction with diverse code rates. Fig. \ref{PRO4} (c) shows the graphical interface at the receiver side, when we transmit the real-time VR video using 64 QAM symbols. In the graphical interface, the important parameters are displayed in real time, including the received spectrum, constellation, signal power, data rate, etc. We can see that the received 64 QAM symbols can be clearly separated, and our prototype supports the real-time transmission of high-definition VR video. Finally, it has been verified that the EIRP of the RIS with 2-bit elements-based wireless communication prototype developed is 51.7 dBm if the transmitted power is 30 dBm, which has a 2 dB gain compared to RIS with 1-bit elements-based wireless communication systems \cite{RRA11}.

Furthermore, as shown in Fig. \ref{PRO5}, we also construct the RIS-based wireless communication prototype at 28.5 GHz. Fig. \ref{PRO5} (a) shows the designed RIS operating at 28.5 GHz, whose measured gain is 19.1 dBi, while the transmitter and receiver are shown in Fig. \ref{PRO5} (b) as well as Fig. \ref{PRO5} (c).

\section{Conclusions}\label{S5}

In this paper, we have proposed, constructed and measured a RIS having 256 2-bit elements, which significantly reduces the power consumption and hardware cost of the conventional phased arrays. Based on the RIS having 256 2-bit elements, we have designed the world's first RIS-based wireless communication prototype for supporting energy-efficient wireless communications. Specifically, the prototype designed consists of modular hardware and flexible software, including the hosts for parameter setting and data exchange, the USRPs for baseband and RF signal processing, as well as the RIS for signal transmission and reception. Our experimental evaluation has demonstrated the feasibility and efficiency of RIS in wireless communications. The test results showed that a 21.7 dBi antenna gain can be obtained by the proposed RIS at 2.3 GHz, while at 28.5 GHz, a 19.1 dBi antenna gain can be achieved. Furthermore, the OTA test results showed that the proposed RIS-based wireless communication system significantly reduces the power consumption without degrading the performance in terms of EIRP compared to existing wireless communication systems. Our prototype will find a wide range of applications in the near future, such as wireless communications in complex terrains (e.g., mountains, snowfields, deserts and offshore areas), high-speed air-to-ground and air-to-air data transmission, deep space communication, near-earth satellite communication, mobile hotspot coverage, etc.


\end{document}